# Spin Hall magnetoresistance at Pt/CoFe$_2$O$_4$ interfaces and texture effects

Miren Isasa[1], Amilcar Bedoya-Pinto[1], Saül Vélez[1], Federico Golmar[1,2,3], Florencio Sánchez[4], Luis E. Hueso[1,5], Josep Fontcuberta[4] and Fèlix Casanova[1,5]

[1]CIC nanoGUNE, 20018 Donostia-San Sebastian, Basque Country, Spain.
[2]I.N.T.I.-CONICET, Av. Gral. Paz 5445, Ed. 42, B1650JKA, San Martín, Bs As, Argentina.
[3]ECyT, UNSAM, Martín de Irigoyen 3100, B1650JKA, San Martín, Bs. As., Argentina
[4]Institut de Ciència de Materials de Barcelona (ICMAB-CSIC), Campus UAB, 08193 Bellaterra, Catalonia, Spain.
[5]IKERBASQUE, Basque Foundation for Science, 48011 Bilbao, Basque Country, Spain.

**Abstract**
We report magnetoresistance measurements on thin Pt bars grown on epitaxial (001) and (111) CoFe$_2$O$_4$ (CFO) ferrimagnetic insulating films. The results can be described in terms of the recently discovered spin Hall magnetoresistance (SMR). The magnitude of the SMR depends on the interface preparation conditions, being optimal when the Pt/CFO samples are prepared *in situ,* in a single process. The spin-mixing interface conductance, the key parameter governing SMR and other relevant spin-dependent phenomena such as spin pumping or spin Seebeck effect, is found to be different depending on the crystallographic orientation of CFO, highlighting the role of the composition and density of magnetic ions at the interface on spin mixing.

Spintronics exploits the spin-dependent charge transport in solids. Pure spin currents, in which spin angular momentum with no electric charge is transported, is expected to lead to a new generation of faster and low-energy consumption spintronic devices [1]. Several methods to create pure spin currents have been developed in the recent years, including non-local spin injection [2,3,4], spin pumping [5,6,7], direct spin Hall effect (SHE) [7,8] or spin Seebeck effect [9,10,11,12]. The detection of these pure spin currents can be done via the inverse spin Hall effect (ISHE) [8,13]. Platinum is the most commonly used non-magnetic metal (NM) for spin current to charge current conversion [6,7,8,9,11,14].

Spin currents, in the form of spin wave excitations, can propagate in ferromagnetic insulators (FMI), for long distances. NM/FMI bilayers are used to create (*via* SHE) and/or detect them (*via* ISHE) [11,14]. Within this framework, a new type of magnetoresistance, so called "spin Hall magnetoresistance" (SMR), has been recently discovered in Pt/YIG [15-20]. As sketched in Fig. 1, SMR arises from the simultaneous effect of SHE and ISHE in the NM (Fig. 1a), combined with the presence of a FMI in one of the interfaces. The generated spin current can be absorbed by the magnetization **M** as a spin-transfer torque when **M** is perpendicular to **s** (Fig. 1b), where **s** is the spin polarization, or reflected when **M** and **s** are parallel (Fig. 1c). Therefore, the charge current in the NM layer varies and its resistance will depend on the magnetization direction at the surface of the FMI. So far, SMR has only been reported for NM/FMI being FMI soft ferromagnets such as YIG [16-19], and more recently Fe$_3$O$_4$ and NiFe$_2$O$_4$ [20].

The concept of spin-mixing conductance [21], which determines the efficiency of the spin current transport at the interface, is at the base not only of SMR, but also of spin Seebeck effect and spin pumping [22]. The nature of the NM/FMI interface strongly affects the observation of such phenomena [17,23,24,25,26]. A detailed comprehension of the mechanisms behind the spin-mixing conductance concept is thus important for a better understanding and control of all these spin-dependent effects. Instrumental for the purpose of this research, we select $CoFe_2O_4$ (CFO), a room-temperature ferrimagnetic insulating oxide [27]. The presence of $Co^{2+}$ ions anticipates a large magnetic anisotropy in CFO [28] and the competing nature of magnetic interactions in spinels may lead to different magnetic properties [29] at (001) and (111) surfaces. Therefore, CFO is especially suitable to explore the role of the surface magnetic textures by using SMR. In this work, we report magnetoresistance measurements on Pt layers grown on (001) or (111) epitaxial CFO films, displaying features fully compatible with SMR, with different spin-mixing conductances for (001) and (111) interfaces. This observation is in agreement with recent speculations that spin-mixing conductance anisotropy in ferrimagnetic spinels could be larger than in YIG [21, 30].

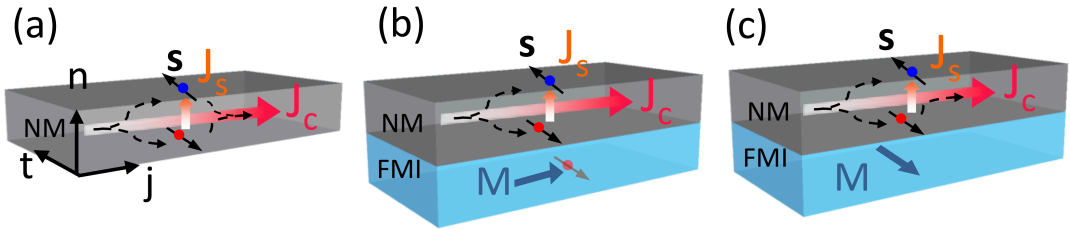

**Figure 1.** *(a) NM layer with strong spin-orbit coupling, with a charge current $j_c$ flowing along **j**. A spin current $j_s$ along **n** with spin polarization **s** along **t** is created due to SHE. The spin current is reflected back at the surfaces, generating additional charge current due to ISHE. (b) NM/FMI bilayer where the magnetization in the FMI is perpendicular to the spin polarization of the spin current. In this case, the spin current will be absorbed at the NM/FMI interface. (c) When the magnetization is parallel to the spin polarization, the spin current will be reflected. The difference in resistance between (b) and (c) leads to SMR.*

CFO films were grown on (001) and (111) $SrTiO_3$ (STO) substrates. The deposition was carried using a CFO stoichiometric target by pulsed laser deposition using a KrF laser with fluence of 1.5(3) $J/cm^2$ and a repetition rate of 5 Hz at a temperature of about 550 °C and oxygen pressure $P_{O2}$= 0.1 mbar [31]. The thickness of the CFO films ranged from 40 nm to 67 nm (see Table I), as inferred from growth rate calibration by X-ray reflectometry [27]. A total of five pairs of Pt/CFO samples were prepared by using two substrate orientations: STO(001) and STO(111), and three distinct processes denoted: EX-1, EX-2 and IN (Table I). In samples prepared by processes EX-2 and IN, the CFO layers were grown simultaneously on (001) and (111) substrates in each run, whereas the Pt layer, deposited by dc sputtering, was grown either ex-situ (EX-2) or in-situ (IN). In process EX-1, the CFO layers on (001) and (111) substrates were grown in different runs and the Pt layer ex-situ. For EX-1 and EX-2 samples, the thickness of Pt was kept constant (around 7 nm). In case of IN samples, Pt of different thicknesses were grown (6.5, 4 and 2 nm). The Pt layers in the ex-situ processes EX-1 and EX-2 were deposited at room temperature whereas in the in-situ IN process, the Pt was grown at 400º C. For the transport measurements the Pt layers were patterned into Hall bars (width W=100 μm and length L=800 μm), as sketched in Fig. 2. For EX-1 and EX-2, patterning was done by using electron-beam lithography with positive resist followed by dc sputtering of the Pt and lift-off, fabricated on top of the CFO films. The Pt layers of IN samples were patterned by using electron-beam lithography with negative resist followed by Ar-ion milling and resist removal. A sample of Pt/YIG was also grown for control experiments using a commercial (111) YIG films. Magnetotransport measurements were performed at 300 K in a cryostat with external magnetic fields (**H**) ranging from -9 T to 9 T applied at different angles. Two different configurations, longitudinal and transverse (see sketches in Fig. 2), have been used for the electrical measurements.

The presence of SMR is assessed by performing angle-dependent magnetoresistance (ADMR) measurements. In Fig. 2 we show, as illustrative examples, the longitudinal and transverse ADMR measured for (001)EX-1(7) and (111)EX-2(7) samples, measured at 9 T, in three relevant **H**-rotation planes defined in sketches of Fig. 2. Baseline resistances of $R_{L0}$=338 Ω (Figs. 2a-c) and $R_{L0}$=763 Ω (Figs. 2e-g) for the longitudinal configuration and $R_{T0}$=24.7 mΩ (Fig. 2d) and $R_{T0}$=824 mΩ (Fig. 2h) for the transverse configuration have been subtracted for clarity. According to the current understanding of SMR [15,16,20], the angular dependence of the longitudinal resistivity $\rho_L$ and the transverse resistivity $\rho_T$ measured in the NM layer are given by:

$$\rho_L = \rho_0 + \rho_1(1 - m_t^2) \qquad (1)$$

$$\rho_T = \rho_2 m_n + \rho_3 m_j m_t \qquad (2)$$

where **m**($m_j$, $m_t$, $m_n$)=**M**/$M_s$ are the cosine directors of the magnetization **M** along the **j**-, **t**- and **n**-directions; $M_s$ is the saturation magnetization of CFO; $\rho_0$ is the baseline resistivity of the NM layer; $\rho_1/\rho_0$ is the SMR; and $\rho_2$ accounts for an anomalous Hall-like contribution. According to this theoretical model, $\rho_3 = \rho_1$ [15,16,20]. As the measurements shown in Fig. 2 have been performed at fields (9 T) much larger than the coercive field $H_C$ of the CFO film and where the film-magnetization is reversible [27], we assume that **m** roughly follows **H**, i.e. **m**∥**H**.

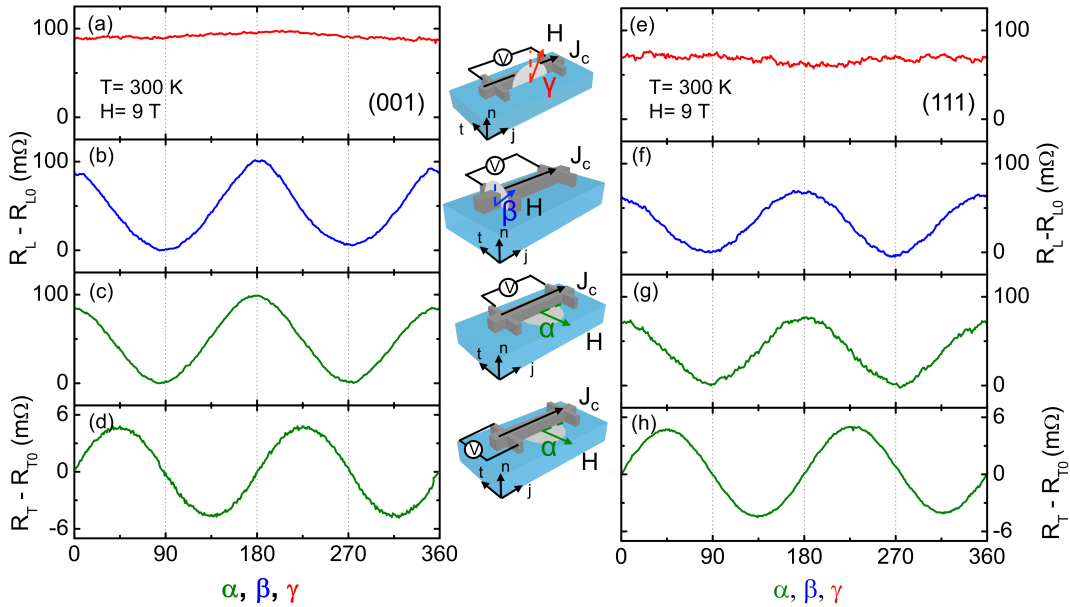

**Figure 2.** *Angle-dependent magnetoresistance measurements at 9 T and 300 K for (001)EX-1(7) **(a-d)** and (111)EX-2(7) **(e-h)** samples. $R_{L,T}$ is the measured resistance and $R_{L0,T0}$ is the subtracted background. (a-c and e-g) Longitudinal resistance $R_L$ as a function of the direction of the applied magnetic field, in three different rotation planes. (d and h) Transverse*

*resistance $R_T$ as a function of angle α. Central panel: sketches indicate the definition of the angles α, β, γ and the measurement configuration.*

The longitudinal resistance $R_L(\gamma)$ (Figs. 2a and 2e) does not show any angular dependence, therefore an anisotropic magnetoresistance (AMR≈$\cos^2\gamma$) [32] of Pt, induced by proximity effect [33] of the neighboring ferromagnetic CFO layer, is excluded. In contrast, a constant $R_L(\gamma)$ is in agreement with Eq. 1. $R_L(\beta)$, plotted in Fig. 2b and 2f, can be described by $R_L(\beta)\approx\cos^2\beta$. This dependence agrees also with the SMR prediction (Eq. 1) (with $\rho_1 > 0$). Similarly, $R_L(\alpha)$ data shown in Fig. 2c and 2g can also be described by $R_L(\alpha)\approx\cos^2\alpha$. In this configuration, both AMR [32] and SMR (Eq. 1) might contribute but, as argued above, AMR has been found to be negligible and thus the observed α-dependence can be safely ascribed to SMR. The transverse resistance $R_T(\alpha)$, shown in Fig. 2d and 2h, displays a $\cos\alpha\times\sin\alpha$ dependence, fully consistent with Eq. 2. In summary, the observed ADMR response of the (001)EX-1(7) and (111)EX-2(7) samples indicates the prevalence of SMR in Pt/CFO with both epitaxial (001) and (111) CFO textures.

The amplitude of the angular variation of the longitudinal resistance for the (001)EX-1(7) sample is $\Delta R_L$=90 mΩ and thus SMR is $\rho_1/\rho_0=\Delta R_L/R_{L0}=2.7\times10^{-4}$. The change in the transverse resistance ($\Delta R_T$=9.22 mΩ) is smaller than $\Delta R_L$ by ~10, in agreement with the difference on the geometrical factor (L/W~8), and yields the expected $\rho_1=\rho_3$ relation [15,16,20]. The magnitude of SMR is given by [15,16]:

$$\frac{\rho_1}{\rho_0} \approx \theta_{SH,NM}^2 \frac{\frac{2\lambda_{NM}^2}{\sigma_{NM}t_{NM}}G_r\tanh^2\frac{t_{NM}}{2\lambda_{NM}}}{1+\frac{2\lambda_{NM}}{\sigma_{NM}}G_r\coth\frac{t_{NM}}{\lambda_{NM}}}, \qquad (3)$$

where $\sigma_{NM}$, $\lambda_{NM}$, $\theta_{SH,NM}$ and $t_{NM}$ are the conductivity, spin diffusion length, spin Hall angle and thickness of the NM element (Pt), respectively, and $G_r$ is the real part of the spin-mixing conductance at the Pt/CFO interface. $G_r$ governs the spin transfer torque at the interface and thus the efficiency of spin injection [21,23,26]. $G_r$ can be extracted from the magnitude of SMR using Eq. 3, if the other parameters are known. The discrepancy in the values of $\theta_{SH,Pt}$ and $\lambda_{Pt}$ existing in the literature [34] has been clarified very recently [35] and, accordingly, we will use the values $\theta_{SH,Pt}$ = 0.056 and $\lambda_{Pt}$ = 3.4 nm given in Ref. [35]. For the (001)EX-1(7) sample, we get $G_r$ = 2.4×10$^{14}$ Ω$^{-1}$m$^{-2}$, which is similar to values reported in literature for other NM/FMI systems; indeed, for Pt/YIG, it ranges from 1.2×10$^{12}$ to 1.3×10$^{15}$ Ω$^{-1}$m$^{-2}$ [14,16,17,18,19,20,22,24,26,36], 1.9×10$^{14}$ Ω$^{-1}$m$^{-2}$ for Au/YIG [23] or 2.6×10$^{14}$ Ω$^{-1}$m$^{-2}$ for Pt/Fe$_3$O$_4$ [37]. A detailed comparison between our results and previous works is difficult due to the use of different set of $\theta_{SH,Pt}$, $\lambda_{Pt}$ parameters for the calculation of $G_r$ and different fabrication conditions. A more direct comparison could be done with the magnitude of SMR: our result lies within the range of values from 1.9×10$^{-4}$ to 9.5×10$^{-4}$ obtained for Pt/YIG with similar Pt thicknesses [16,17,20], but also in this case the different fabrication conditions seem to influence SMR value. For example, a control experiment in a Pt/YIG sample fabricated by the same EX-1 process gives us $\rho_1/\rho_0 = \Delta R_L/R_{L0} = 0.7 \times 10^{-4}$ [27]. Of higher interest for the purpose of this paper, however, is the comparison between the $G_r$ of Pt/(001)CFO and Pt/(111)CFO samples. It can be observed in Figs. 2f and 2g that, for the (111)EX-2(7) sample, the change in longitudinal resistance ($\Delta R_L = 69$ mΩ) and the spin Hall magnetoresistance term $\rho_1/\rho_0 = \Delta R_L/R_{L0} = 0.9 \times 10^{-4}$ are smaller than for the (001)EX-1(7) sample. This leads to a smaller $G_r = 2.4 \times 10^{13}$ $\Omega^{-1}$m$^{-2}$.

This result suggests that $G_r$ depends on the relevant crystallographic planes [(001) vs (111)] forming the Pt/CFO interface. Before proceeding with the analysis of this experimental observation, we show in Table I the spin Hall magnetoresistance, at 9 T and 300 K, and the extracted $G_r$ values for all samples, in which we have used the same set of parameters ($\theta_{SH,Pt} = 0.056$ and $\lambda_{Pt} = 3.4$ nm [35]). We will first focus on the samples with Pt thickness of ~7 nm. Inspection of data in Table I immediately reveals some remarkable trends: (i) For all pair of 7-nm-thick Pt samples (IN, EX-2 and EX-1), $G_r(001)$ is different and somewhat larger than the corresponding $G_r(111)$ and (ii), although the CFO layers have been grown under nominally identical conditions in samples EX-2 and IN, the extracted spin-mixing conductance differs, being definitely larger for IN than for EX-2 samples. Regarding (ii), it is well known that $G_r$ is very sensitive to the details of the interface between the FMI and the NM [17,23,24,25,26]. As the Pt layer is deposited differently in EX-2 and IN samples (ex-situ and in-situ, respectively) the interface is likely modified during the ex-situ Pt deposition, because it involves exposure of the free surface of the CFO to air and to the chemicals used for the lithography process. Consequently, it is not surprising to find a larger $G_r$ value for IN than for EX-2 samples and therefore $G_r(001)$ and $G_r(111)$ values for IN samples set upper bounds to the spin-mixing conductances of (001) and (111) interfaces in Pt/CFO. Regarding (i), the systematic observation that for every pair of samples $G_r(001) > G_r(111)$ suggests that the spin-mixing conductance may depend on the interface orientation of the ferromagnetic insulator.

| Sample | Fabrication process | Crystal. orient. | $t_{CFO}$ (nm) | $t_{Pt}$ (nm) | $\rho_0$ (μΩcm) | $\Delta R_L/R_{L0}$ | $G_r$ ($\Omega^{-1}$m$^{-2}$) |
|---|---|---|---|---|---|---|---|
| (001) EX-1(7) | EX-1 | (001) | 67 | 7 | 29.6 | $2.7 \times 10^{-4}$ | $2.4 \times 10^{14}$ |
| (111) EX-1(7) | EX-1 | (111) | 56 | 7 | 19.5 | $0.2 \times 10^{-4}$ | $1.4 \times 10^{13}$ |
| (001) EX-2(7) | EX-2 | (001) | 57 | 7 | 29.7 | $1.2 \times 10^{-4}$ | $7.4 \times 10^{13}$ |
| (111) EX-2(7) | EX-2 | (111) | 57 | 7 | 66.8 | $0.9 \times 10^{-4}$ | $2.4 \times 10^{13}$ |
| (001) IN(7) | IN | (001) | 40 | 6.5 | 21.4 | $2.5 \times 10^{-4}$ | $2.4 \times 10^{14}$ |
| (111) IN(7) | IN | (111) | 40 | 6.5 | 18.2 | $1.8 \times 10^{-4}$ | $1.9 \times 10^{14}$ |
| (001) IN(4) | IN | (001) | 40 | 4 | 20.2 | $3.4 \times 10^{-4}$ | $2.6 \times 10^{14}$ |
| (111) IN(4) | IN | (111) | 40 | 4 | 23.3 | $2.5 \times 10^{-4}$ | $1.4 \times 10^{14}$ |
| (001) IN(2) | IN | (001) | 40 | 2 | 36.0 | $6.0 \times 10^{-4}$ | $2.4 \times 10^{14}$ |
| (111) IN(2) | IN | (111) | 40 | 2 | 34.5 | $4.3 \times 10^{-4}$ | $1.1 \times 10^{14}$ |

**Table I**. *Summary of relevant data corresponding to the five pairs of Pt/CFO samples used in this work: fabrication process, crystallographic orientation of CFO film, thickness of CFO film ($t_{CFO}$), thickness of Pt film ($t_{Pt}$), resistivity of the Pt film ($\rho_0$), SMR effect ($\rho_1/\rho_0 = \Delta R_L/R_{L0}$) and the real part of the spin-mixing conductance ($G_r$) calculated from Eq. 3 by using $\theta_{SH,Pt} = 0.056$ and $\lambda_{Pt} = 3.4$ nm [35].*

From this analysis, where pairs of samples prepared using different fabrication processes are compared, we infer that SMR is a robust phenomenon that is present in Pt/CFO, although a quantitative comparison between crystallographic orientations can be best done for IN samples due to the optimal interface preparation conditions. For this reason, we will now focus on the samples prepared with the same IN process and different Pt thicknesses: (001)IN(2,4,7) and (111)IN(2,4,7). We show in Fig. 3 the dependence of the magnetoresistance, at 9 T and 300 K, of the three pairs of IN samples when rotating the magnetic field in a plane perpendicular to the current (i.e. as a function of $\beta$). In this geometry, the amplitude of the observed magnetoresistance ($\rho_1/\rho_0 = \Delta R_L/R_{L0}$) is linked to $G_r$ (Eq. 3) and it thus allows us a simple visualization of the changes of $G_r$ and its evaluation. It can be appreciated in Table I that the extracted $G_r$ values for these samples are radically different for both terminations [$G_r(001) = 2.5(1) \times 10^{14}$ $\Omega^{-1}m^{-2}$ and $G_r(111) = 1.5(4) \times 10^{14}$ $\Omega^{-1}m^{-2}$] and largely independent of the Pt thickness when considering the same crystallographic orientation. This last observation, which is expected as $G_r$ is basically an interfacial property, demonstrates the good reproducibility of the Pt/CFO interface achieved in our fabrication IN process. Therefore, the $G_r$ values are consistently different between orientations [$G_r(001) > G_r(111)$ for any Pt thickness], being a solid evidence of the anisotropy of the spin-mixing conductance.

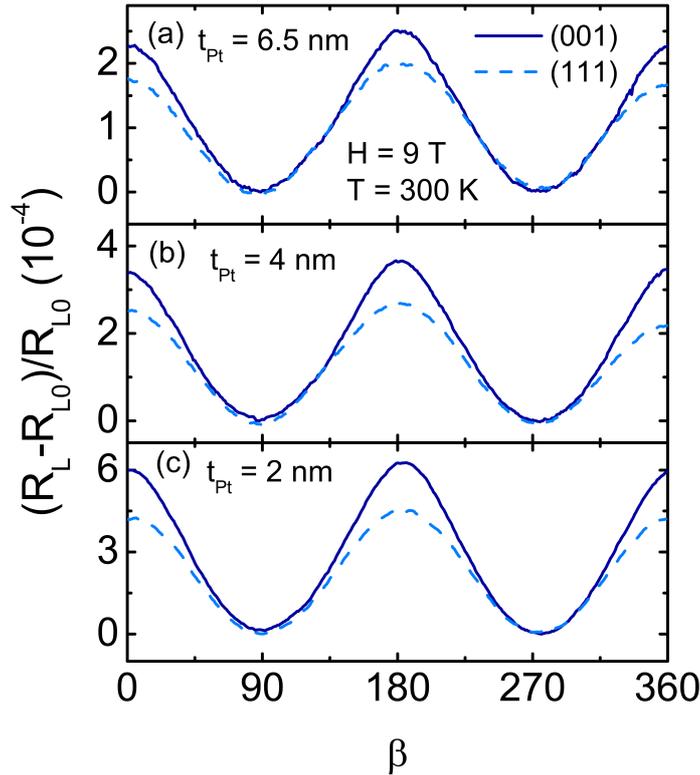

**Figure 3**. *Angle-dependent longitudinal magnetoresistance, at 9 T and 300 K, when rotating the magnetic field in a plane perpendicular to the current (i.e. as a function of β) for epitaxial (001) and (111) CFO/Pt samples grown in situ with (a) 6.5 nm, (b) 4 nm and (c) 2 nm of Pt. $R_L$ is the measured resistance and $R_{L0}$ is the subtracted background.*

Since the density of magnetic ions at the interface and their magnetic orientation determine the spin transfer, any detailed understanding for the observed difference $G_r(001) > G_r(111)$ should start by considering the microscopic nature of the atomic planes involved at the interface. This is far from obvious in spinel $AB_2O_4$ oxides; for instance in (111) there are 6 different atomic planes all of them being polar and, therefore, unstable. There are different mechanisms to solve this dipole-associated electrostatic energy divergence and, for this reason, the surface termination in (001) and (111) planes of spinel oxides is strongly dependent on the conditions used to prepare the surfaces. As a result, a definitive conclusion is still missing even for the most studied case of $Fe_3O_4$ (see Ref. 38 for a recent review). Nevertheless, theoretical and experimental trends indicate that in (001) surfaces the termination containing tetrahedrally coordinated $Fe^{3+}$ ions is most commonly found, whereas in (111) surfaces both oxygen and tetrahedral terminations are more favorable [38]. A similar situation has been suggested for $MgAl_2O_4$ [39] and $CoFe_2O_4$ [40].

Recent first-principles calculations of $G_r$ for different surfaces of $CoFe_2O_4$ [41] predict values of $2.82 \times 10^{14} \, \Omega^{-1} m^{-2}$ for the tetrahedral termination in the case of (001) orientation and $0.63 \, (1.15) \times 10^{14} \, \Omega^{-1} m^{-2}$ for the oxygen (tetrahedral) terminations in (111) orientation. The values for these stable (111) terminations are smaller than that predicted for the most stable (001) termination, which are similar to our experimental values and in agreement with the higher stability of the tetrahedrally coordinated $Fe^{3+}$ planes in (001) and tetrahedrally coordinated $Fe^{3+}$ and oxygen-terminated planes in (111) as argued above.

To conclude, we have shown that spin Hall magnetoresistance is at the origin of the longitudinal and transverse magnetoresistance of Pt films deposited on epitaxial (001) and (111) ferrimagnetic insulating CFO thin films. Although the observed SMR is a robust phenomenon, its magnitude depends on the interface preparation conditions, being optimal when the samples are prepared *in situ*. The spin-mixing conductance at Pt/CFO is found to be similar to those reported for other NM/FMI heterostructures. Most importantly, the observation that (001) and (111) CFO films have clearly different SMR illustrates that atomic configuration of the magnetic atoms at NM/FMI interfaces have an important effect in the spin-mixing conductance, a crucial parameter which is also at the base of other relevant spin-dependent phenomena, such as spin pumping or spin Seebeck effect. These results might have important implications for the design of future spintronic devices based on insulators.

**Acknowledgments**

This work is supported by the European Union 7th Framework Programme under the Marie Curie Actions (PIRG06-GA-2009-256470-ITAMOSCINOM), NMP project (NMP3-SL-2011-263104-HINTS), and the European Research Council (Grant 257654-SPINTROS), by the Spanish Ministry of Science and Education (MAT2012-37638, MAT2011-29269-C03 and CSD2007-00041), by the Basque Government (PI2011-1) and Catalan Government (2009 SGR 00376). M. I. acknowledges the Basque Government for a PhD fellowship (BFI-2011-106). J. F. acknowledges stimulating discussions with Xavier Martí.

**Spin Hall magnetoresistance at Pt/CoFe$_2$O$_4$ interfaces and texture effects**

Miren Isasa[1], Amilcar Bedoya-Pinto[1], Saül Vélez[1], Federico Golmar[1,2,3], Florencio Sánchez[4], Luis E. Hueso[1,5], Josep Fontcuberta[4] and Fèlix Casanova[1,5]

[1]CIC nanoGUNE, 20018 Donostia-San Sebastian, Basque Country, Spain.
[2]I.N.T.I.-CONICET, Av. Gral. Paz 5445, Ed. 42, B1650JKA, San Martín, Bs As, Argentina.
[3]ECyT, UNSAM, Martín de Irigoyen 3100, B1650JKA, San Martín, Bs. As., Argentina
[4]Institut de Ciència de Materials de Barcelona (ICMAB-CSIC), Campus UAB, 08193 Bellaterra, Catalonia, Spain.
[5]IKERBASQUE, Basque Foundation for Science, 48011 Bilbao, Basque Country, Spain.

**SUPPLEMENTARY INFORMATION**

### I. Description of CoFe$_2$O$_4$ (CFO)

We select CoFe$_2$O$_4$ (CFO), a room-temperature ferrimagnetic insulating oxide, for the present study. It has a cubic spinel structure (A)[B$_2$]O$_4$, where A and B indicate tetrahedrally and octahedrally coordinated sites. In the ideal inverse structure the Fe$^{3+}$ ions are equally distributed among A and B sites, whereas Co$^{2+}$ ions are confined to B sites, i.e. (Fe$^{3+}$)[Co$^{2+}$Fe$^{3+}$]O$_4$. The strong antiferromagnetic interaction between ions at A and B sublattices determine the ferrimagnetic ordering. In general, however, some partial degree of inversion occurs and CFO is better described as (Fe$_{1-x}$Co$_x$)[Fe$_{1+x}$Co$_{1-x}$]O$_4$.

### II. Experimental details of CFO growth

CFO films were grown on (001) and (111) SrTiO$_3$ (STO) substrates. The deposition was carried using a CFO stoichiometric target by pulsed laser deposition using a KrF laser with fluence of 1.5(3) J/cm$^2$ and a repetition rate of 5 Hz at a temperature of about 550 °C and oxygen pressure P$_{O2}$= 0.1 mbar [S1]. The thickness of the CFO films ranged from 40 nm to 67 nm, as inferred from growth rate calibration by X-ray reflectometry. The STO substrate has a cubic perovskite structure with cell parameter a$_{STO}$ = 3.905 Å. Bulk CFO is also cubic (a$_{CFO}$ = 8.392 Å). The structural mismatch between the film and substrates (f=+6.9%) would impose a biaxial compressive in-plain stress on the CFO films. X-ray, θ/2θ scans using a Siemens D-5000D diffractometer and Cu-Kα$_{1,2}$ radiation, were used to confirm that all films were fully out-of-plane textured without spurious phase. The positions of the (002) or (111) reflections of the substrate were used for internal angular calibration. The out-of-plane cell parameter was found to be d$_{(001)}$ = 8.392 Å and d$_{(111)}$ = 4.839 Å for (001)- and (111)-oriented CFO films, respectively, thus indicating that all films, independently on the thickness and orientation, are virtually relaxed, as expected from the large structural mismatch. The surface roughness was determined by atomic force microscopy; it was found that the roughness of all films was of about 0.2 - 0.3 nm.

### III. Magnetic characterization of (001) and (111) CFO films

Magnetization measurements were performed at 100 K and 300 K in the same cryostat as the magnetotransport measurements with external magnetic fields (**H**) ranging from -9 T to 9 T applied at different angles. A vibrating sample magnetometer (VSM) was used to determine the magnetization of the CFO films.

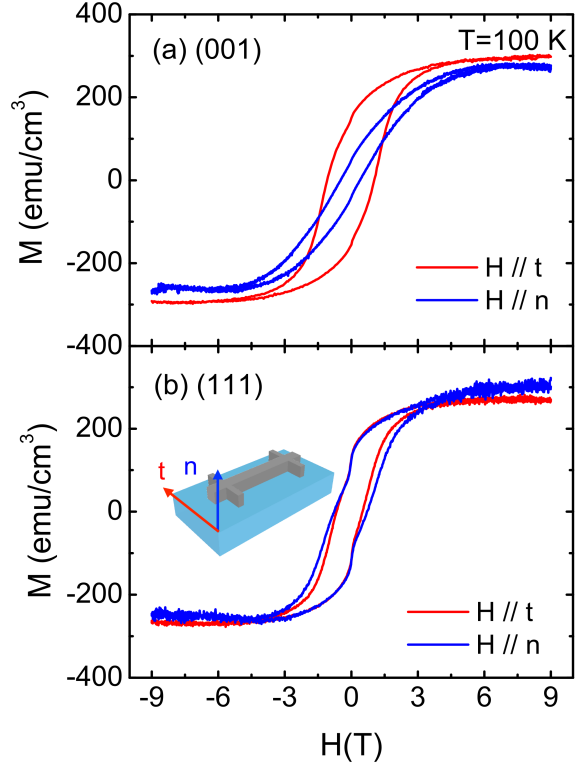

**FIG. S1.** *Magnetic hysteresis loops for the (001)EX-1(7) sample (a) and (111)EX-2(7) sample (b) in the cases in which H is applied along **t** (red curve) and **n** (blue curve), as defined in the inset.*

Figure S1a shows the hysteresis loops M(H) of the (001)EX-1(7) sample at 100 K obtained by applying the magnetic field **H** along **t** and **n** directions. The magnetization curve with **H** along **j** is not shown because it is indistinguishable from the one along **t**, indicating in-plane magnetic isotropic behavior. The large coercive fields $H_c(\mathbf{j},\mathbf{t}) \approx \pm 1.06$ T and the fact that hysteresis only disappears at $\approx$ 5-6 T are signatures of the strong magnetic anisotropy typical of CFO thin films [S2,S3]. The shape of the hysteresis loop when the field is applied out-of-plane indicates a harder magnetization axis and, correspondingly, the coercive field $H_c(\mathbf{n}) \approx \pm 0.44$ T and the magnetic remanence are smaller. The saturation magnetization ($M_s$=290 emu/cm$^3$) is lower than the corresponding bulk value as commonly observed in spinel thin films [S4-S7] and attributed to the presence of antiphase boundaries (APB) [S4,S5] or to surface anisotropy effects [S1]. The diamagnetic background, arising mainly from the STO substrate, has been corrected by subtracting a linear term $\chi_d H$, where $\chi_d$ is the high-field slope of the raw data. The $\chi_d$ values are practically identical for all **H** orientations, as expected for the cubic STO substrate (not shown). Note that the presence of such background, however, would conceal any possible contribution from non-saturating behavior of the CFO film at high fields, as commonly observed in these systems [S1,S2,S4,S6,S8].

The magnetization loops for (111)EX-2(7) sample are shown in Fig. S1b. They share common features with those of the (001) CFO film of Fig. S2a, such as the saturation magnetization ($M_s$=290 emu/cm$^3$) and the isotropy of the in-plane magnetization (not shown). A distinct feature is that, whereas in the (001) sample the out-of-plane direction is a harder magnetization axis than in-plane, in the (111) sample it is an easier magnetization axis. This change in the magnetic anisotropy of (001) and (111) CFO films has already been reported [S2].

## IV. Control samples: Pt/SiO$_2$ and Pt/YIG

Platinum Hall bars with the same geometry (width W=100 μm and length L=800 μm) as the samples used for the main study were fabricated on top of silicon oxide (SiO$_2$) and yttrium iron garnet (YIG). Thermally grown 150-nm-thick SiO$_2$ on Si substrate was used for the first case and 3.5-μm-thick YIG grown by liquid phase epitaxy on a (111) gadolinium gallium garnet (GGG) single crystal substrate was used for the second case. The Pt Hall bars were fabricated using electron-beam lithography with positive resist followed by dc sputtering of the Pt and lift-off, following the same process and equipment as the batches of samples EX-1 and EX-2 used in the main text.

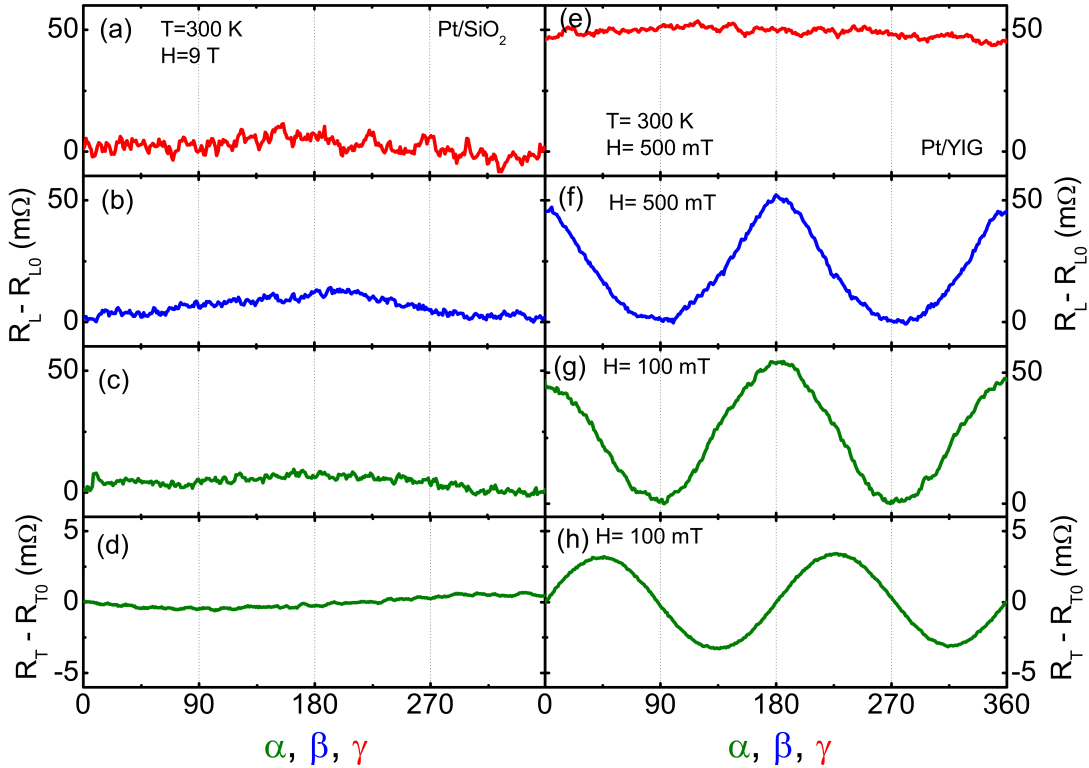

**FIG. S2.** *Angle-dependent magnetoresistance measurements at 300 K and 9 T for Pt/SiO$_2$ (a-d) and at 300 K and different magnetic fields for Pt/YIG (e-h) samples. $R_{L,T}$ is the measured resistance and $R_{L0,T0}$ is the subtracted background. (a-c and e-g) Longitudinal resistance $R_L$ as a function of the direction of the applied magnetic field, in three different rotation planes (angles α, β and γ). (d and h) Transverse resistance $R_T$ as a function of angle α.*

Angle-dependent magnetoresistance (ADMR) with longitudinal and transverse configuration measurements at 300 K are shown in Fig. S2, in three relevant **H**-rotation planes, as defined in the main text (angles α, β and γ). For the Pt/SiO$_2$ sample, no variation of resistance was observed for any angular variation up to 9 T (Figs. S2a-d), which demonstrates that no artifact is present in the measurement. For the Pt/YIG sample, the ADMR confirms the occurrence of spin Hall magnetoresistance (SMR), showing that a FMI is needed in order to observe the effect (Figs. S2e-h): $R_L(\gamma)$ does not show any variation, $R_L(\beta)$ and $R_L(\alpha)$ show the same amplitude ($\Delta R_L$=50 mΩ) and angular dependence ($\cos^2$) and $R_T(\alpha)$ shows a $\sin\alpha\cdot\cos\alpha$ dependence and an amplitude ($\Delta R_T$=6.6 mΩ) that follows $\Delta R_L/\Delta R_T \approx L/W$. Different values of the magnetic field were used in the various ADMR measurements for the Pt/YIG sample. $R_L(\alpha)$ and $R_T(\alpha)$ are plotted at 100 mT, because the in-plane magnetization of YIG is saturated. $R_L(\beta)$ and $R_L(\gamma)$ are measured at 500 mT, because the out-of-plane magnetization saturates at a higher field due to the shape anisotropy of the thin film.

The SMR value ($\rho_1/\rho_0=\Delta R_L/R_{L0}=0.7\times10^{-4}$) obtained from the Pt/YIG sample yields a spin mixing conductance of $G_r=4.1\times10^{13}\,\Omega^{-1}m^{-2}$ (using $\theta_{SH,Pt}= 0.056$ and $\lambda_{Pt}= 3.4$ nm [S9]), which is in agreement with recent values obtained from SMR in Pt/YIG bilayers [S10-S13].